\documentclass[conference]{IEEEtran}
\IEEEoverridecommandlockouts
\usepackage{cite}
\usepackage{url}
\usepackage{hyperref}
\usepackage{amsmath,amssymb,amsfonts}
\usepackage{algorithmic}
\usepackage{graphicx}
\usepackage{textcomp}
\usepackage{xcolor}
\def\BibTeX{{\rm B\kern-.05em{\sc i\kern-.025em b}\kern-.08em
    T\kern-.1667em\lower.7ex\hbox{E}\kern-.125emX}}
\begin{document}

\title{Towards Better Adaptive Systems by Combining MAPE, Control Theory, and Machine Learning 
}

\author{\IEEEauthorblockN{Danny Weyns}
\IEEEauthorblockA{
\textit{KU Leuven, Belgium}\\
Linnaeus University, Sweden \\
danny.weyns@kuleuven.be}
\and
\IEEEauthorblockN{Bradley Schmerl}
\IEEEauthorblockA{
\textit{Carnegie Mellon University}\\
Pittsburgh, USA \\
schmerl@cs.cmu.edu}
\and
\IEEEauthorblockN{Masako Kishida}
\IEEEauthorblockA{
\textit{National Institute of Informatics}\\%
Tokyo, Japan \\
kishida@nii.ac.jp}
\and
\IEEEauthorblockN{Alberto Leva}
\IEEEauthorblockA{
\textit{Politecnico di Milano}\\
Milan, Italy \\
alberto.leva@polimi.it}
\and
\hspace{40pt}
\IEEEauthorblockN{Marin Litoiu}

\IEEEauthorblockA{
\hspace{35pt}
\textit{York University}\\
\hspace{40pt}
York, Canada \\
\hspace{40pt}
mlitoiu@yorku.ca}
\and
\IEEEauthorblockN{Necmiye Ozay}
\IEEEauthorblockA{
\textit{University of Michigan}\\
Ann Arbor, MI, USA \\
necmiye@umich.edu}
\and
\IEEEauthorblockN{Colin Paterson}
\IEEEauthorblockA{
\textit{University of York}\\
York, United Kingdom \\
colin.paterson@york.ac.uk}
\and
\IEEEauthorblockN{Kenji Tei}
\IEEEauthorblockA{
\textit{Waseda University}\\
Tokyo, Japan \\
ktei@aoni.waseda.jp}
}

\maketitle

\begin{abstract}
Two established approaches to engineer adaptive systems are architecture-based adaptation that uses a  Monitor-Analysis-Planning-Executing (MAPE) loop that reasons over architectural models (aka Knowledge) to make adaptation decisions, and control-based adaptation that relies on principles of control theory (CT) to realize adaptation. 
Recently, we also observe a rapidly growing interest in applying machine learning (ML) to support different adaptation mechanisms. 
While MAPE and CT have particular characteristics and strengths to be applied independently, in this paper, we are concerned with the question of how these approaches are related with one another and whether combining them and supporting them with ML can produce better adaptive systems. %
We motivate the combined use of different adaptation approaches using a scenario of a cloud-based enterprise system and illustrate the analysis when combining the different approaches. To conclude, we offer a set of open questions for further research in this interesting area.  
\end{abstract}


\section{Introduction}

Designing self-adaptive systems with some level of autonomy has been studied for over two decades~\cite{Oriezy1999,1516538,Cheng2009,deLemos2013,978-3-319-74183-3_1,WeynsBook2020}. A number of different approaches have been used to engineer such systems. Two established approaches are (i) architecture-based adaptation that relies on Monitor-Analysis-Planning-Executing (MAPE) components that reason over architectural models (aka Knowledge) to make adaptation decisions~\cite{Kephart,Rainbow,Kramer2007,Weyns2012-1,weyns2017software}, and (ii) control-based adaptation that relies on principles of control theory (CT) to realize adaptation of a target system~\cite{Hellerstein:2004,Filieri:2017,Shevtsov:2018}. 
Recently, we also observe a rapid growing interest in applying machine learning (ML) to support different adaptation mechanisms~\cite{gheibi2021applying}, for instance to use classifiers to reduce large adaptation spaces~\cite{quin2019efficient}, learn a model of the system with MAPE~\cite{Bencomo19}, or learn adaptation rules with CT~\cite{7515437}. 

This paper is concerned with the following question: 
\vspace{3pt}\\
\textit{How can MAPE and CT be combined with and supported by ML techniques to produce better adaptive systems?}  
\vspace{3pt}\\
%
%
In answer to this question, this paper contributes insights into: (1) the characteristics and strengths of both CT and MAPE, and how ML can support adaptation techniques, (2)~how CT and MAPE can be combined with and supported by ML, 
and (3) a number of open topics for further research in this area. 
%
%
This paper is an outcome of a working group of the 3$^{rd}$ Shonan Meeting on Controlled Adaptation of Self-adaptive Systems.\footnote{The 3$^{rd}$ Shonan Meeting on Controlled Adaptation of Self-adaptive Systems (CASaS). Report: \url{https://shonan.nii.ac.jp/docs/No.153.pdf}}

The remainder of this paper is structured as follows.  In Section~\ref{section:related-work} we summarise a selection of related work. Section~\ref{section:definitions-scope}  introduces the main concepts and outlines the scope of this work. Section~\ref{section:approaches} highlights strengths of CT and MAPE and looks into support of ML to adaptation techniques. Section~\ref{section:scenarios} zooms in on combining CT and MAPE and supporting them with ML in an example case. Finally, Section~\ref{section:conclusions} wraps up and outlines a set of open questions for further research. 

\section{Related Work}\label{section:related-work}

We summarise a selection of related work, starting with approaches that combine MAPE with CT. Then we look at work that combines MAPE or CT with ML. We conclude by motivating the research question posed in this paper. 
\vspace{3pt}\\
\textbf{Combining MAPE with CT.} 
DYNAMICO is a conceptual model that 
considers three levels of dynamics in adaptive systems that map to three interacting feedback loops~\cite{TamuraVMDS13}.  
A first feedback loop monitors requirements (adaptation goals) to ensure their fulfilment. A second feedback loop manages context information preserving context information relevant to adaptation. Finally, a third feedback loop controls the target system according to control objectives, while taking into account the context. The feedback loops can be realised using different adaptation mechanisms, e.g., MAPE for the first and second and CT for the third feedback loop.  
%
The idea of combining MAPE-based discrete decision-makers/planners with continuous low-level controllers has also been proposed in the context of symbolic control, with applications in robotics~\cite{belta2007symbolic}, self-driving cars and driver-assist systems~\cite{wongpiromsarn2012receding,nilsson2015correct}. In these approaches, the higher levels guarantee satisfaction of user goals (often specified in temporal logics) under discrete external factors, while lower levels ensure robust tracking of the goals set by the higher levels under disturbances.
%
%
\vspace{3pt}\\
\textbf{Supporting MAPE with ML.} 
As the complexity of self-adaptive software has steadily  grown over time, the opportunities to exploit ML to support MAPE have also increased. 
Just as ML has found 
substantial success in recent years in the perception pipeline of automotive vehicles, such techniques have been deployed to monitor computing systems at runtime to derive meaningful measures from data gathered in complex environments~\cite{metzger2019feature} and to detect faults~\cite{affonso2015framework}. During the analysis phase, ML has been used when the space of possible adaptations grows too large to be handled with traditional techniques~\cite{DoncktWQDM20} or where patterns may be extracted from large data sets (e.g., network traffic~\cite{Maimo2018}). Finally, we see ML used in the planning stage where selecting a policy for adaptation becomes difficult due to the size of the adaptation space. Here ML can pre-filter adaptation options and present a set of Pareto optimal solutions from which the MAPE solution can choose an appropriate action~\cite{Jamshidi2019}, or evolutionary search can be used to change control parameters to more optimal settings~\cite{Caldas2020}.
%
\vspace{3pt}\\
\textbf{Supporting CT with ML.} ML has been used in control theoretical solutions for a long time. One established area is the use of ML for model and system identification, see for instance~\cite{ASTROM1971123,annurev-control-053018-023744,SJOBERG1994359,047134608X.W1046.pub2}. Recent examples that use ML techniques to derive a linear model of a system are~\cite{2568225.2568272,3328730}, while~\cite{wu2019machine} uses recurrent neural networks to capture the behaviour of non-linear systems.
More generally, neural networks have been used for a long time to solve highly non-linear control problems~\cite{nguyen1990neural,chow1998recurrent}.  Recently, reinforcement learning has gained increasing attention to deal with control problems where the state space is large and unknown a priori~\cite{Kamalapurkar2016}, or the number of controller parameters that need to be tuned is large~\cite{wen2019online}.
%
\vspace{3pt}\\
\textbf{Motivation for Research Question.} 
While some efforts show benefits of combining MAPE with CT on the one hand and the potential of exploiting ML to support the adaptation mechanisms on the other hand, further research is required on how these approaches can be combined to build better adaptive systems. In particular, further research is required to understand: (1) the characteristics and strengths of CT and MAPE and the use of ML for adaptation, (2) how CT and MAPE can be combined with and supported by ML, (3) how we can consolidate this knowledge into reusable assets. 

\section{Concepts and Scope}\label{section:definitions-scope}

We briefly describe the main concepts and the scope of the work presented in this paper. We start with the two adaptation techniques we focus on: CT and MAPE, and conclude with ML that we aim to use in support for adaptation. 

\subsection{Control Theory}


In this paper, for CT we essentially mean classical control structures -- single-loop or at most feed forward compensation, cascade controls -- made of individually simple linear time-invariant blocks such as Proportional-Integral-Derivative controllers (PID), running at a fixed sampling rate or triggered by events. These structures have demonstrated to be powerful at keeping some variable(s) at prescribed set points or within prescribed ranges in the face of disturbances, provided that the relationship between the control signal and the controlled variable is not too complex and does not vary too much over time~\cite{LevaEtAl2013}. In large applications, the ``classical CT layer'' is often extended with a higher ``advanced layer'' realised as Model Predictive Control (MPC). This upper layer feeds commands to the lower one, adapting its behaviour when the conditions of the system (plant) require such an action.


This two-layer control scheme fits domains like process control very well, where the plant is ruled by clearly defined physical laws, but in general does not seem to fit equally well adaptive software. Among the reasons are that similar laws are not so easy to define for software systems, that objectives can be very heterogeneous, and time scales for decision-making may not fit well~\cite{Hellerstein:2004,Shevtsov:2018}. Moreover, adaptive software systems often require complex types of adaptations, e.g., changing the structure of a managed system. 
%
%
\vspace{3pt}\\
\noindent\textbf{Key insights}: CT's strength lies in keeping variables at prescribed set points or within prescribed ranges, regardless of disturbances. Adding an MPC control layer atop a classical control layer fits well for process control, but does not seem to map naturally to software systems. We conjecture that in adaptive software the two control schemes can be re-used, where the upper layer may be realised using MAPE.

\subsection{MAPE}

MAPE concerns techniques usually covered by the software engineering for self-adaptive systems community. MAPE is an acronym introduced by Kephart and Chess~\cite{Kephart2003} referring to the essential activities (functions) that should be covered by any systems with adaptive capabilities: Monitoring, Analysis, Planning, and Execution. The aim of MAPE is usually to adapt software -- its configuration, structure, behaviour, etc. -- rather than physical phenomena. While all forms of control are likely to have one or more of these activities (or are known in other guises as Sense-Plan-Act, for example), we use MAPE as a label in this paper to capture approaches to adapt software and the techniques that have been developed to handle this.

The MAPE activities are centered around Knowledge models that typically include various forms of runtime models~\cite{Blair2009}, such as software architecture models of the managed system and environment, goal models, Markov networks that allow predicting  qualities of different system configurations, etc. These models usually focus on the properties that one wishes to maintain the software~\cite{Camara2016SMGs,Iftikhar16,3180155.3182540}. Monitoring is used to update these models with data about the current state of the software, the environment in which it is running, and the goals of the software.
Analysis typically involves taking the current state and determining if the properties align with user or business goals. Often, analysis is focused on a multi-dimensional space that trades off different quality attributes of the software -- for example, balancing performance, reliability, security, or cost. Tools that can be used here include runtime simulations, model checking, architecture analysis, etc. Analysis can be expressed as thresholds or constraints that the software should achieve or maintain. E.g., we might desire a response time less than two seconds; we may aim to minimise operational costs.
Planning then decides the set of steps required to change the software to satisfy the properties. In simple condition-based MAPE systems, planning would just be choosing which action to take based on some condition. Other approaches might choose some organised sequence or tree of actions that maximise some utility function. Sometimes more advanced techniques are required, possibly based on AI, to generate plans. 
Finally, execution takes these actions and manages their enactment on the actual software system. This may involve non-trivial synchronisation with the system that is adapted.

A pioneering work that applies different levels of control is IBM's Autonomic Computing Reference Architecture~\cite{ibm2005architectural}, see also~\cite{VILLEGAS201717}. Yet this work considers hierarchies of MAPE loops rather than integrations of MAPE and CT-based solutions. 

A key aspect for the industrial adoption of MAPE is frameworks that offer interfaces for monitoring and safe adaptations of the underlying system. A typical example is OSGi\,\cite{2207947} that offers a dynamic component model for Java. 
\vspace{3pt}\\
\textbf{Key insights}: MAPE addresses adaptation of software rather than physical properties or resources; MAPE has a global perspective on the system (or subsystems being managed); MAPE deals with trading off requirements/quality attributes to determine what needs to be changed about the software.

\subsection{Machine Learning}

ML techniques can be considered in four dimensions~\cite{shalev2014understanding,bishop2006pattern}. First: unsupervised vs supervised vs  interactive.  
    An unsupervised learner aims at finding patterns in data sets without labels. Cluster analysis is an example~\cite{Amparo2013}. A supervised learner learns a function that maps input to output based on example pairs. An interactive learner collects the input-output pairs by interaction with the environment. A classic example is reinforcement learning~\cite{Sutton2018}. Second: active vs passive. An active learner queries some information source in the environment (e.g., a user or teacher~\cite{Settles2009}) to obtain the outputs at new data points that are then used to affect the environment. A passive learner only perceives the information from the environment without affecting it.
Third: adversarial vs non-adversarial. Adversarial learning attempts to fool models through malicious input. An example is using obfuscated spam messages in email filtering~\cite{Blanzieri2008}. Non-adversarial learning has no concept of malicious input.
 Fourth: online vs batch~\cite{93335.93365}. Online learning updates the learning model using sequential data, while batch learning learns the model using the entire training data set at once.
%



We focus here on ML techniques that support building models and/or strategies for CT and MAPE. In applications with high dimensional data that operate in uncertain environments, it is often difficult to manually build models with appropriate precision. ML, especially deep learning, has shown to be a powerful tool to build effective models with small size and high precision by predicting future behaviours based on previously observed data in uncertain environments~\cite{bengio2009learning,schmidhuber2015deep}.

\section{Characteristics of Approaches}\label{section:approaches} 

We outline characteristics of CT and MAPE emphasising strengths, and look at the support ML can offer to adaptation. 

\subsection{Control Theory}
%
\noindent\textbf{Discrete- and continuous-time models}: 
CT works with both discrete-time and continuous-time models~\cite{franklin1998digital}, like those used in ``fluid'' modelling for queuing systems~\cite{kumar2007stochastic}. Continuous-time models are often written based on conservation laws. Their parameters have a natural physical meaning (e.g., the maximum speed of a CPU), and the effect of modifying them can be foreseen based on physical intuition. 
%
\vspace{3pt}\\
\textbf{Preserving system properties}: 
both for the discrete- and the continuous-time cases, CT offers formal means to assess whether the system will not drift away from the desired operating point (stability) and what are the tolerable modifications to the system that preserve a given property, such as settling time (robustness)~\cite{hinrichsen2011mathematical}. As long as linear theory applies~\cite{hespanha2018linear}, which is often obtained by convenient local-in-the-operating-point approximations, such checks can be done offline and do not require measured data or synthetic stimuli.
\vspace{3pt}\\
\textbf{Atomic vs. sequence of control actions}: controllers naturally lend themselves to ``atomic'' control actions, like setting a value for some input to the controlled system. They are not equally apt to more articulated control actions, for example involving a sequence of operations. If some characteristic parameter (e.g., maximum CPU speed) of the system changes, CT can accommodate for such variations as long as a means to detect them online is available (adaptive control~\cite{aastrom2013adaptive}). If the same detection implies complex operations on large sets of data, possibly analysing the history of the system, alternative techniques (ML) may be more appropriate.

\subsection{MAPE}
%
\noindent\textbf{Complex combination of sensor data}: raw data may be of low-level and need to be combined in complex ways to generate meaningful knowledge to reason about adaptations; different dimensions may apply to combine data, for which MAPE offers support, e.g., combining measurements over time, or integrating data from different data sources. 
\vspace{3pt}\\
\textbf{Complex relation between observable data and properties of interest}: deriving properties that are needed to reason about adaptation from observable parameters may be complex and require advanced models and analysis techniques that naturally fit MAPE, e.g., interference in a wireless network may be represented as parameters of a probabilistic model; packet loss can then be predicted using online model checking~\cite{3180155.3182540}. 
\vspace{3pt}\\
\textbf{Complex quality goals need to be combined}: stakeholders often require an adaptive system to provide different quality goals; these goals may be different in nature, e.g., maximise profit, minimise delay, ensure a minimum level of performance, ensure a constant throughput; this leads to conflicts and requires potentially complex trade-off analysis. Such types of analysis are at the heart of MAPE~\cite{Cheng2006}. 
\vspace{3pt}\\
\textbf{Variability}: the variation of a software system under control can be parametric or structural (e.g., adding/removing components, changing their connections, and changing deployment of components); MAPE approaches can handle the structural variability in analysis and planning. 
\vspace{3pt}\\
\textbf{Switching types of adaptations/modes}: MAPE offers means for adaptations ranging from system parameter tuning to architectural reconfigurations; the latter requires discrete changes of the system, such as activation/deactivation of components, and switching the operation mode of the system.

\vspace{3pt}
\noindent
\textbf{Long-time scale}: achieving the objectives of adaptive systems may require reasoning and planning over long time spans during which the conditions of the system or the environment may change significantly requiring complex replanning.  
\vspace{3pt}\\
\textbf{Known complex actuation}: actuation on a software system may require performing a sequence of low-level parametric or structural changes; its execution is not instantaneous generally, but its consequence is predictable. MAPE works with such complex actuation types.

\subsection{Machine Learning}
%
\noindent\textbf{Model building}: 
ML provides methods to build a model for dynamical systems using data, even where  first-principle modelling is not possible. This includes, but is not limited to classical system identification techniques used in CT. Recently popular is to use deep learning to complement a rough first-principle model, in order to add non-linear effects and external disturbances, which are difficult to model.
\vspace{3pt}\\
\textbf{Complexity reduction}: 
ML can be used to reduce the complexity or dimension of a model supporting the design of a CT controller efficiently and effectively. Commonly used ML approaches for model reduction include principal component analysis, singular value decomposition, and auto-encoder.
\vspace{3pt}\\
\textbf{Estimating properties}: 
For MAPE, ML can provide an a-priori estimation of performance where the environment contains uncertainties or where the environment is not directly observable. ML can also be used for clustering the data, which may provide increased understanding of the patterns in the data. Indeed ML can be used to generate models for the analysis phase of the MAPE loop to provide complex non-linear inference where deriving models traditionally are difficult or impossible. Dimensionality reduction may also help  understanding or allow computationally intractable problems to be tackled using traditional software techniques.
\vspace{3pt}\\
\textbf{Optimise policies}: 
In many domains deriving plans for MAPE is non-trivial and ML techniques may help. Where the problem can be specified as an abstract state representation, reinforcement learning may be employed to optimise policies, i.e., a plan of action in each state, which can then be encoded as a plan in  the MAPE loop.
\vspace{3pt}\\
\textbf{Designing control input}: 
(Deep) Reinforcement learning may also be used to design a sequence of control inputs with or without using a model. However, unlike MPC, it is not always straightforward to incorporate hard-constraints.





\section{Combining Adaptation  Techniques}\label{section:scenarios}

With the characteristics and strengths of CT and MAPE in hand, we combine the two and support them with ML. To that end, we use a Cloud-based enterprise system. We also studied a second case of a self-driving car, but due to space limitations, we refer to\,\cite{appendix}. Based on our experiences, we propose a first pattern for combining CT with MAPE supported by ML. This pattern offers an initial reusable asset in this area. 

\subsection{Cloud-based Enterprise System}

Consider a cloud application consisting of logical partitions, e.g., a web application deployed across web, application, and data tiers, or an IoT deployed across fog, edge, and central cloud partitions. Partitions are made  of Virtual Machines (VM) with containers (C) that host the application, e.g., microservices that communicate over links within and across VMs. 

A common high-level goal of such applications is to serve users with high quality services, while minimising the cost. Goals can be expressed in terms of \textit{Service Level Agreements} (\textit{SLA}), \textit{budget constraints}, and \textit{user satisfaction}. 
%
%
%
Goals can be translated into technological metrics, such as \textit{response time} and \textit{cost}, which can be considered controlled outputs.  
%
%
The outputs can be affected by control inputs, such as the number of VMs (\textit{\#vm}), the number of containers (\textit{\#containers}), and the number of threads (\textit{\#threads}), buffer or connection pool sizes.  
%
%
Applications are subject to uncertainties (perturbations). 
\textit{Load}, which is the rate of service requests that arrive at the application, 
can be highly unpredictable, non-linear, and multi-dimensional. Another uncertainty is \textit{cloud interference}, which is the effect other applications running on the cloud have on the application. Cloud interference that can affect the CPU, IO, memory, bandwidth, etc., can be highly non-linear and is usually not directly measurable. 
    	\begin{figure*}[h!]
			\centering
			\includegraphics[width=0.8\textwidth]{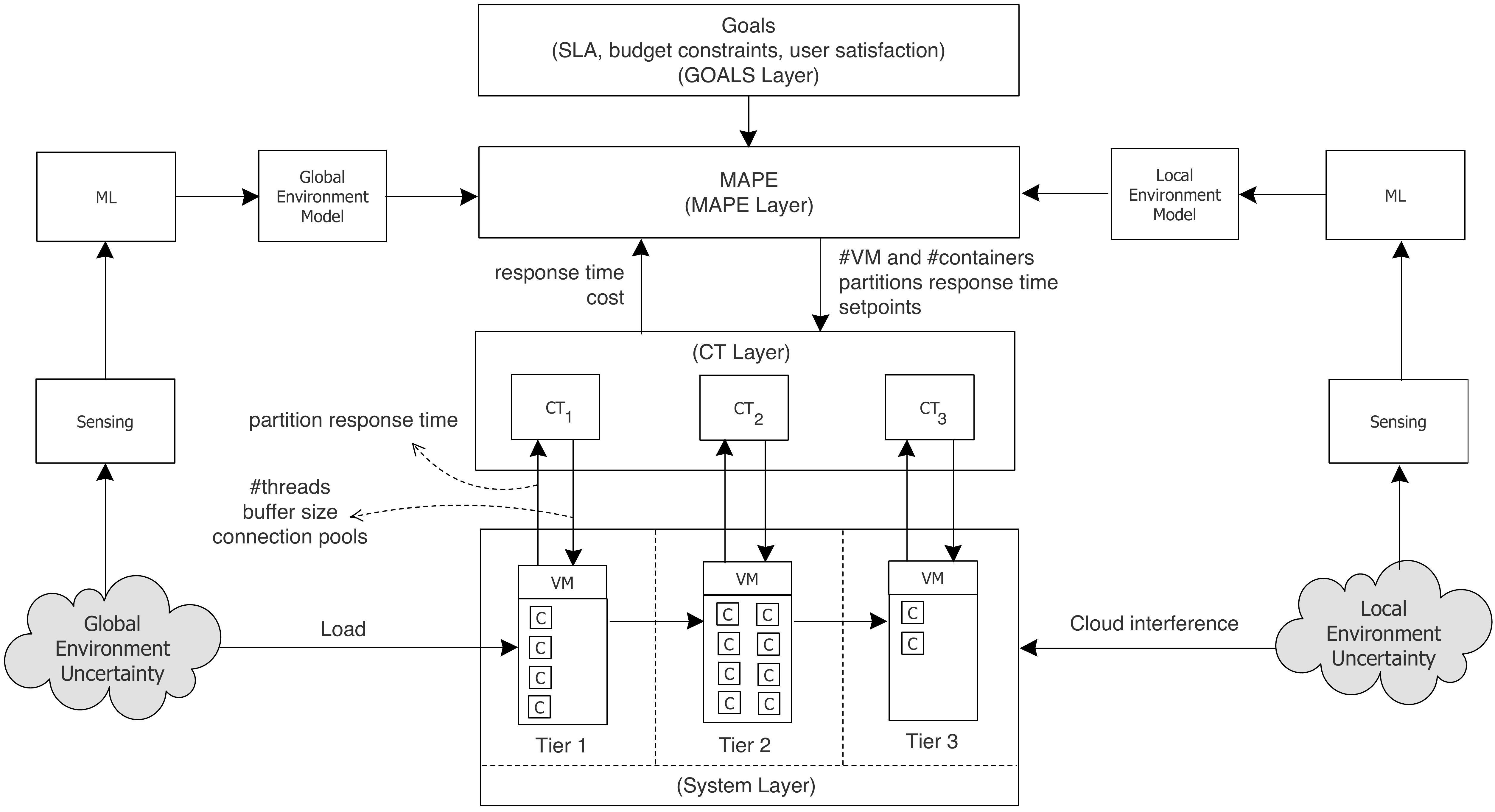}
			\caption{Adaptive Cloud-based enterprise system.} 
			\label{fig:FourTier}
		\end{figure*} 
%
\subsection{Designing the Controller.} 
To achieve the goals of a cloud-based enterprise system in the presence of uncertainties, we combine the strengths of MAPE and CT and support these with ML in a four-layer architecture as shown in Figure~\ref{fig:FourTier}. 
\vspace{3pt}\\
\textbf{Goal layer}: 
The goal layer takes the application owner goals, e.g., an SLA, budget constraints, and user satisfaction, as input and converts them into technological goals for MAPE, such as end-to-end response times and penalties that affect cost.
\vspace{3pt}\\
\textbf{MAPE Layer:}
The MAPE layer monitors the progress of the goals and makes reconfiguration decisions, such as changing the number of VMs/containers on each partition. MAPE will also determine the set points for the lower-level CT controllers. To that end, the MAPE layer monitors the response time of each partition, the end-to-end response time, and the current cost (given by the cost of resources in the cloud eventually the cost model of the cloud provider). To determine the control decisions (number of  VMs/containers and set points for the CT layer), the MAPE layer can rely on techniques such as look-ahead optimisation, search-based algorithms, simulation, or queuing models. The decisions of the MAPE layer are produced in a time scale of order of minutes.
%
%
\vspace{3pt}\\
\textbf{CT layer}: The CT layer controls the number of threads used by the microservices among other local parameters, such as buffer sizes, and connection pools. The decisions of the CT layer are produced in a time scale in the order of milliseconds to seconds. By using control theory-based adaptation, this layer is able to  deal with fast transients and attenuate high frequency uncertainties. CT is appropriate here since the control inputs and their effects on the outputs are close to a continuous time domain, and assurance for stability and robustness is crucial. 
\vspace{3pt}\\
\textbf{System layer}: The system layer that is the subject of adaptation comprises the application logic of a cloud system that is set up as a three-tier architecture as explained above. 
%
\vspace{3pt}\\
\textbf{The role of ML}: ML in our application provides models for: (a) the \textit{load} over time (of the day) as service requests can vary in frequency, data size, and required data outputs, and (b) the \textit{cloud interference} along with high dimensional OS/VM/container metrics over long periods of time. By predicting the request load and expected extra load on the application, the  MAPE layer can make better adaptation decisions.

\subsection{Analysis of Combined Architecture}\label{section:architecture}

\begin{figure}[h!]
			\centering
			\includegraphics[width=0.8\columnwidth]{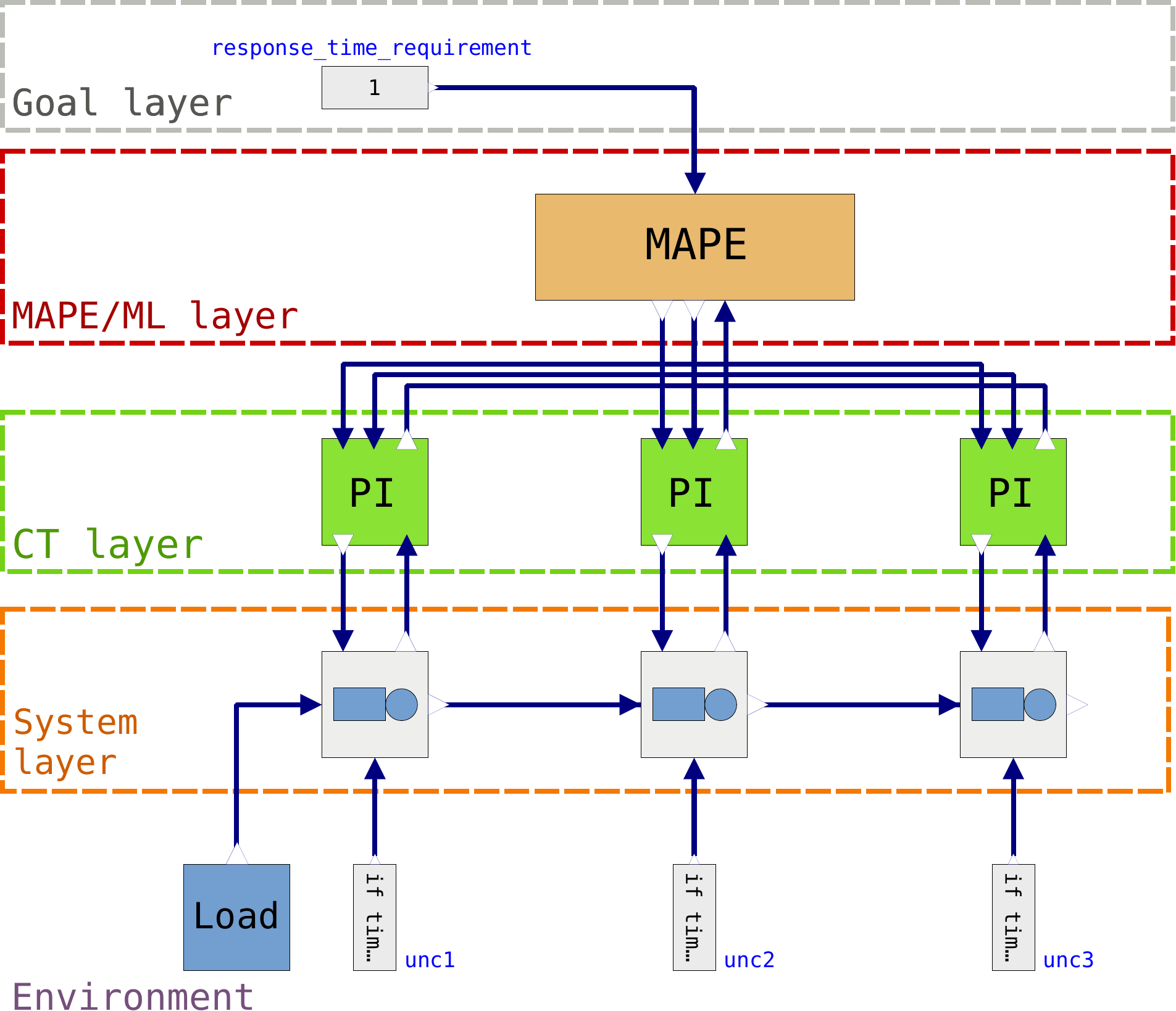}
			\caption{Modelica diagram for a simplified version of the cloud example.}  
			\label{app-sim-diagram}
		\end{figure}

As an initial evaluation of combining MAPE and CT, supported by ML, we model and simulate a simplified version of the three-tier cloud application in Modelica, see Figure~\ref{app-sim-diagram}. 
%
%
The application is modelled in continuous time as three queue-plus-server blocks, each one receiving as input the output of its predecessor. The processing speed is obtained by allotting more or less ``Computational Units'' (CUs for short, e.g., \#threads, \#connections) that can range from 0 to some maximum. By changing that maximum, for example by adding or removing threads, we can re-configuring the system.  

The CT layer comprises  Proportional-Integral (PI) controller that allots CUs to each tier to maintain the desired response time in the face of dynamics of the load and network disturbances (uncertainties \textit{unc}) that affect the throughput. 
%
In addition, the PI controllers compute the number of CUs that they would allow to comply with the required response time if an infinite number of CUs would be available as follows: 
\begin{equation}
 CU_{need} = \frac{CU_{desired}-CU_{max avail}}{CU_{max avail}}
\end{equation}
The CU needs are fed to the upper MAPE/ML layer. Knowing these indices and the required response time for the overall system, this layer can (i) decide the response times for the individual tiers (set points for the PIs) and (ii) reconfigure the system by modifying the maximum CU availability.


	\begin{figure}[htb]
			\centering
			\includegraphics[width=0.95\columnwidth]{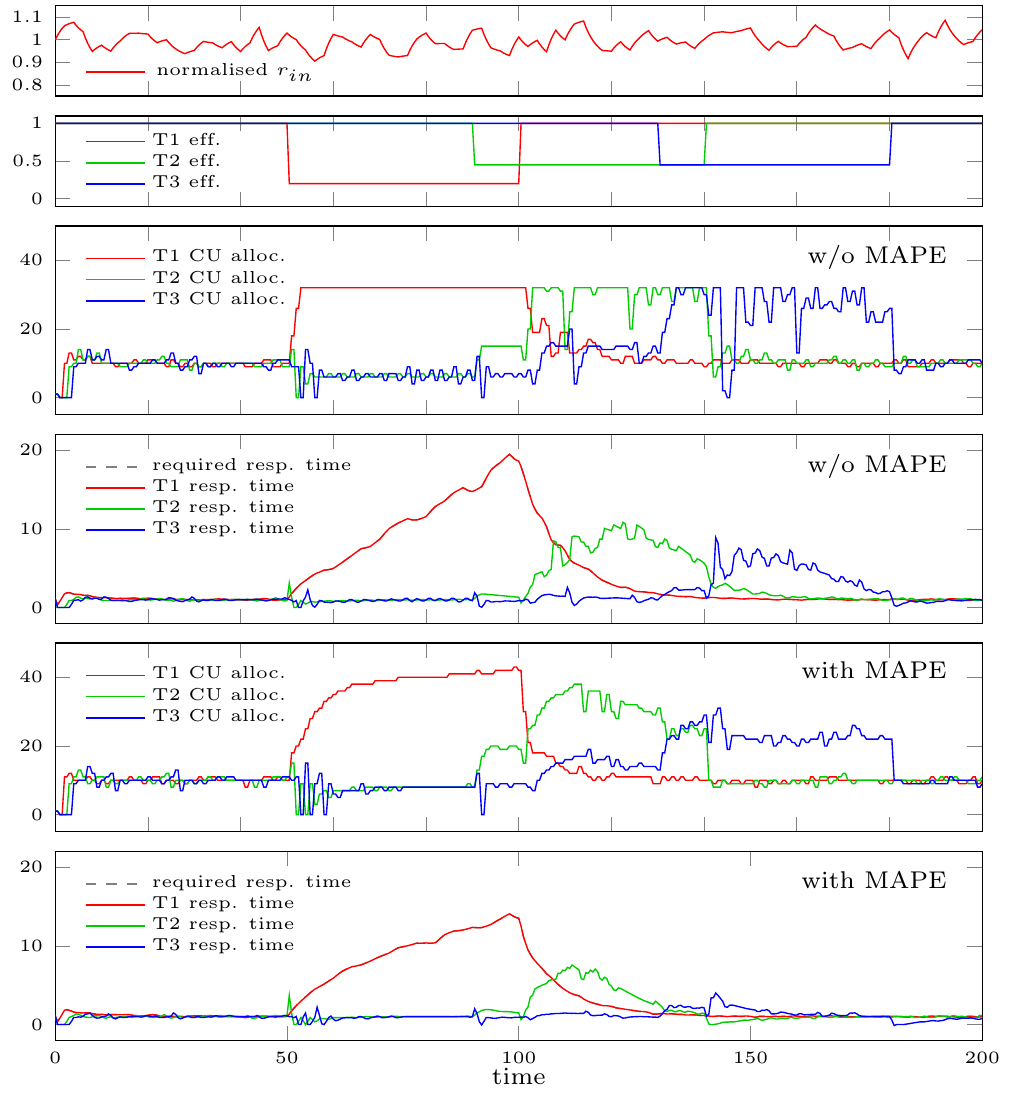}
			\caption{Simulation results: from the top: in-out rate, efficiencies, required and obtained response times, CU allocation, and resource need indices.} 
			\label{app-sim-results}
		\end{figure} 

Figure~\ref{app-sim-results} shows a simulation run with a randomly varying load (input rate $r_{in}$ at the top, plotted normalised to its nominal value) and efficiencies (2nd plot). We see that the CT layer can keep the system on track when this is feasible. We also see that the CU 
need indices do signal the need for a reconfiguration, and more importantly, that their shape distinguishes short-time infeasibilities from sustained ones. 
Dealing with the complex information of these  indices and adapting the system accordingly, is a task for which MAPE/ML is particularly well suited. 
This is illustrated by comparing the CU allocation (3rd and 5th plot) and the obtained response times (4th and 6th plot) respectively without and with activating the functionality of the MAPE/ML layer. The results show that the adaptations applied by the MAPE/ML layer enable the underlying CT layer to better manage temporary infeasibilities.


\subsection{Pattern: 1-MAPE-n-CT with ML for Uncertainty Modelling}\label{section:pattern}  

\begin{figure}[htb]
			\centering
			\includegraphics[width=0.95\columnwidth]{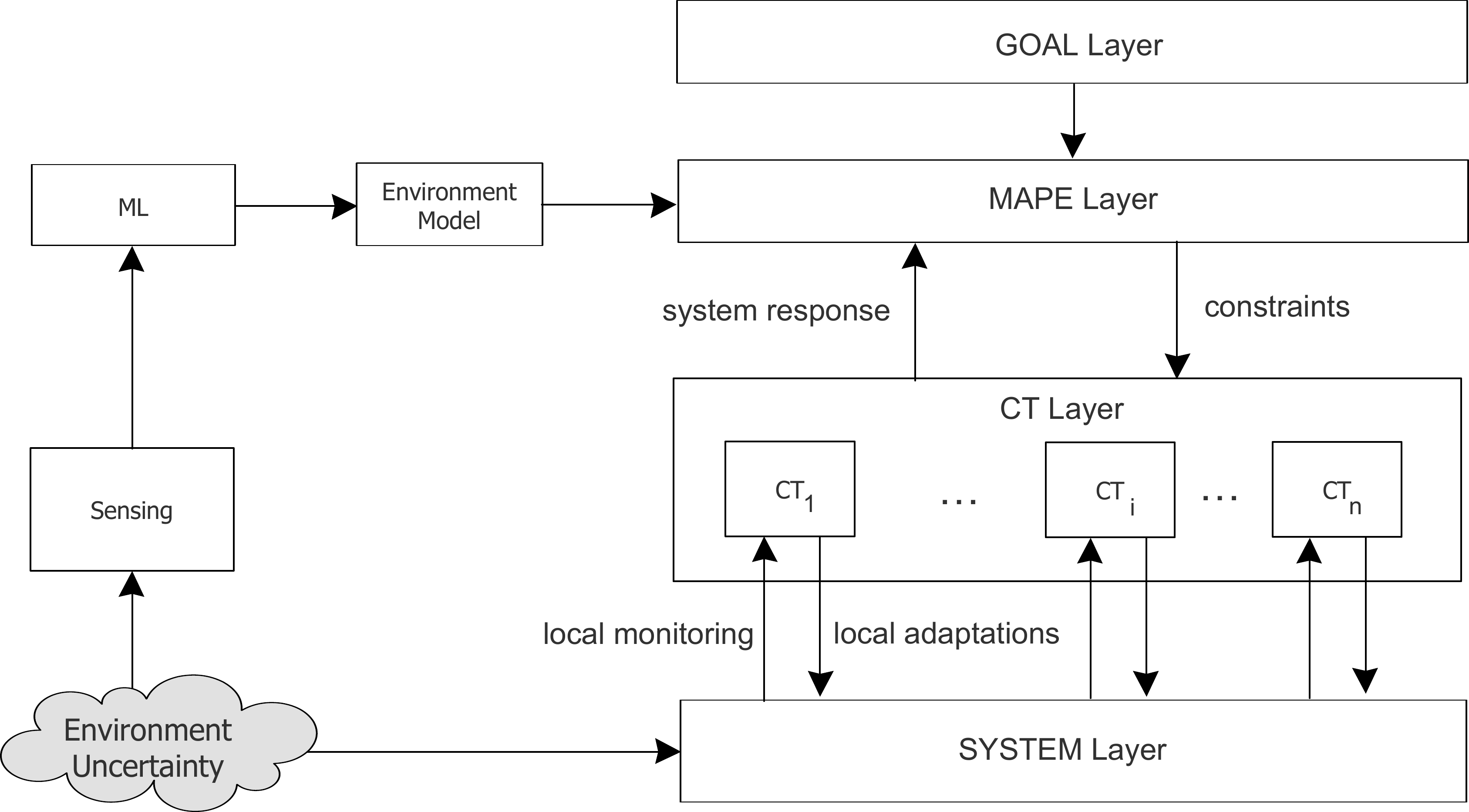}
			\caption{1-MAPE-n-CT with ML for uncertainty modelling pattern.} 
			\label{fig:pattern}
		\end{figure}

Based on our experiences, we identified a first pattern for combining MAPE and CT as shown in Figure~\ref{fig:pattern}.   

The pattern structures the self-adaptive system in four layers where a single MAPE loop is combined with multiple controllers to adapt a distributed system. The rationale for applying the pattern is: (i) CT handles small perturbations, (ii) when CT cannot cope, MAPE is called to adapt the system and change the CT settings to achieve the high-level goals. ML is used to generate up-to-date models of the environment that produces uncertainties the system needs to withstand.

\section{Conclusions and Further Research}\label{section:conclusions} 

This paper presented work in progress on how to combine MAPE and CT and support them with ML to produce better adaptive systems. We characterised the different adaptation approaches: MAPE's strength is its global perspective on the system and its ability to deal with complex trade-offs between requirements. CT's strength lies in keeping variables at prescribed set points or within prescribed ranges, regardless of disturbances. ML, on the other hand, can support MAPE and CT in different ways; one key way is to build runtime models and adaptation strategies from complex and high dimensional data obtained from uncertain environments. 

Second, we illustrated how CT and MAPE can be combined and supported by ML using a use case from the cloud domain. In particular, the case shows how self-adaptive software can be organised in four layers. At the top, a goal layer translates user requirements to operational goals for the adaptation logic. Next, a MAPE layer monitors the progress of the goals, the resources available to the system, and system-wide uncertainties in the environment to determine bounds for the reconfiguration decisions. Then, a CT layer tracks the system behaviour and local uncertainties in the environment, to make reconfiguration decisions of the system within the bounds defined by MAPE. Finally, the system layer comprises the managed system that is subject to adaptation. ML techniques can support the MAPE and CT layers in different functions, for instance, building global and local models of the environment.  
 

While these initial insights are promising, further research is needed to better understand how to combine adaptation approaches. Crucial questions to be answered will be: What are the typical use cases for combining MAPE and CT? How to allocate adaptation responsibilities when MAPE and CT are combined? Can we identify patterns to combine MAPE and CT, and what are their tradeoffs? Can we identify typical coordination mechanisms for MAPE and CT to interact? What are the interesting use cases for ML to be applied in adaptive systems? How can we provide guarantees for the adaptation goals in hybrid architectures that combine CT and MAPE? Answering these questions will require a substantial joint effort among researchers with backgrounds in architecture-based adaptation, control theory, and machine learning. 


\bibliographystyle{IEEEtran}
\bibliography{bib}

\end{document}